\documentclass{article}
\usepackage{spconf,amsmath,graphicx,hyperref, booktabs,balance}

\title{MoEScore: Mixture-of-Experts-based Text-Audio Relevance Score Prediction For Text-to-Audio System Evaluation}

\name{Bochao Sun$^{1}$ \qquad Yang Xiao$^{2}$ \qquad Han Yin$^{3\ast}$}
\address{
    $^{1}$School of Marine Science and Technology, Northwestern Polytechnical University, Xi'an, China \\
    $^{2}$School of Computing and Information Systems, The University of Melbourne, Melbourne, Australia \\
    $^{3}$School of Electrical Engineering, KAIST, Daejeon, Republic of Korea \\
    $\ast$Corresponding author. E-mail: \textit{hanyin@kaist.ac.kr}
}

\begin{document}
\ninept
\maketitle
\begin{abstract}
Recent advances in generative models have enabled modern Text-to-Audio (TTA) systems to synthesize audio with high perceptual quality. However, TTA systems often struggle to maintain semantic consistency with the input text, leading to mismatches in sound events, temporal structures, or contextual relationships. Evaluating semantic fidelity in TTA remains a significant challenge. Traditional methods primarily rely on subjective human listening tests, which is time-consuming. To solve this, we propose an objective evaluator based on a Mixture of Experts (MoE) architecture with Sequential Cross-Attention (SeqCoAttn). 
Our model achieves the first rank in the XACLE Challenge, with an SRCC of 0.6402 (\textbf{an improvement of 30.6\% over the challenge baseline}) on the test dataset. Code is available at: https://github.com/S-Orion/MOESCORE.
\end{abstract}

\section{Introduction}
\label{sec:intro}

Text-to-audio (TTA)~\cite{liu2023audioldm,liu2024audiolcm,huang2023make2} aims to synthesize audio samples from textual descriptions, which has been widely applied in various applications, such as voice assistant~\cite{ren2019fastspeech}, media content production~\cite{arik2017deep, huang2023make}, and virtual reality acoustic environment simulation~\cite{1,yin2025envsdd}. 

Evaluating TTA systems~\cite{wang2025tta} typically involves two core dimensions: audio quality and text-audio relevance. 
While modern TTA models have achieved remarkable progress in generating high-quality audio, they often fail to fully reflect the content of input text, leading to a critical gap in text-audio relevance performance~\cite{lee2024challenge}. Text-audio relevance is conventionally assessed through subjective human evaluation, which is costly, inconsistent, and difficult to scale, motivating the need for objective evaluation metrics.


To address this challenge, the XACLE Challenge 2026\footnote{XACLE Challenge 2026: \url{https://xacle.org/}} aims to develop models that predict text-audio relevance scores (TARS) for TTA systems.
The baseline approach relies on a single cross-modal model and struggles to jointly capture global semantic alignment, fine-grained semantics, and temporal correspondence.
To overcome these limitations, we propose a TARS predictor based on a Mixture-of-Experts (MoE)~\cite{masoudnia2014mixture} architecture enhanced with Sequential Cross-Attention (SeqCoAttn). 

Rather than relying on a single model, we observe that CLAP-family models exhibit complementary inductive biases in text-audio relevance prediction due to differences in granularity modeling, temporal sensitivity, and pretraining objectives.
Motivated by this observation, we integrate multiple specialized experts, including three CLAP-based variants and an expert enhanced with SeqCoAttn, via a feature-aware gating network. The SeqCoAttn-based expert models fine-grained temporal correspondence between audio and text, complementing embedding-based global alignment.
Experiments on the XACLE Challenge 2026 dataset demonstrate that the proposed MoE framework achieves state-of-the-art performance.

\section{PROPOSED METHODS}

\subsection{Model Architecture}

\begin{figure}[t]
\begin{minipage}[b]{1.0\linewidth}
  \centering
  \centerline{\includegraphics[width=0.9\linewidth]{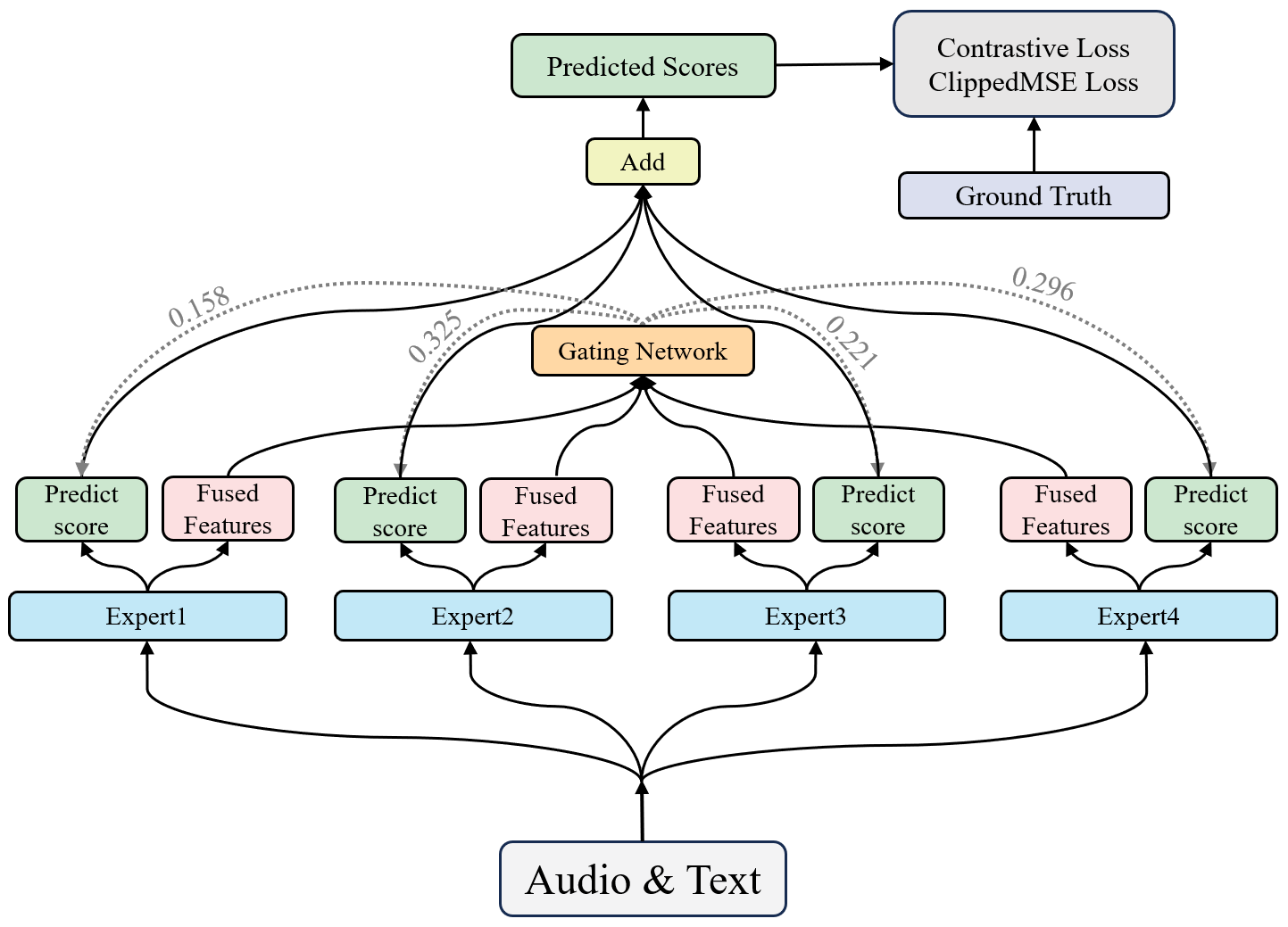}}
  \centerline{(a) Overall Architecture}\medskip
\end{minipage}
\begin{minipage}[b]{1.0\linewidth}
  \centering
  \centerline{\includegraphics[width=0.9\linewidth]{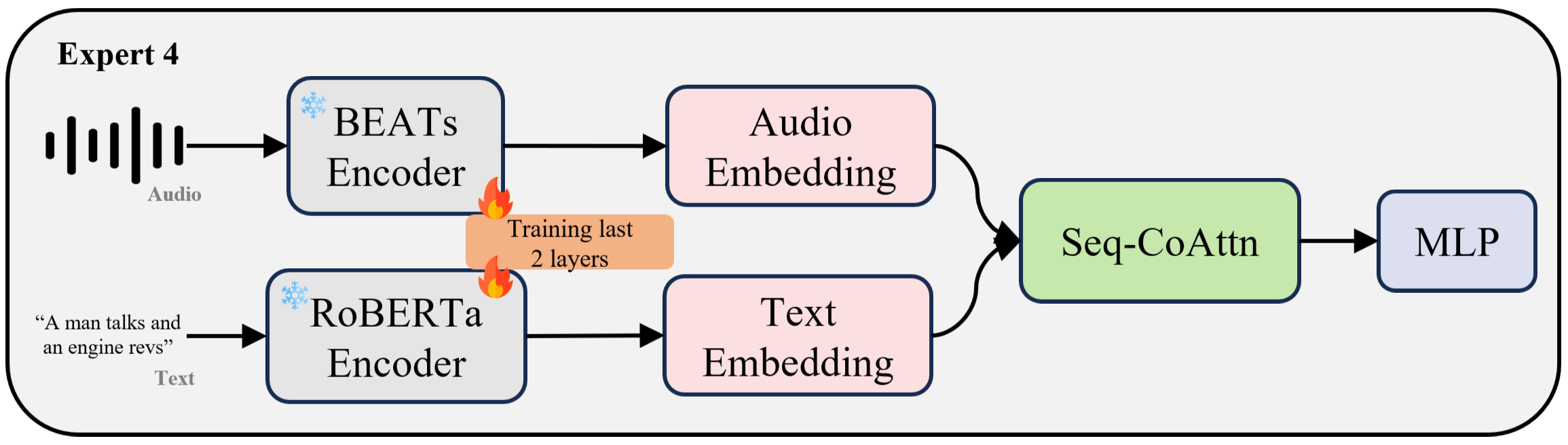}}
  \centerline{(b) Expert-4 Architecture}\medskip
\end{minipage}
\hfill
\begin{minipage}[b]{1.0\linewidth}
  \centering
  \centerline{\includegraphics[width=0.9\linewidth]{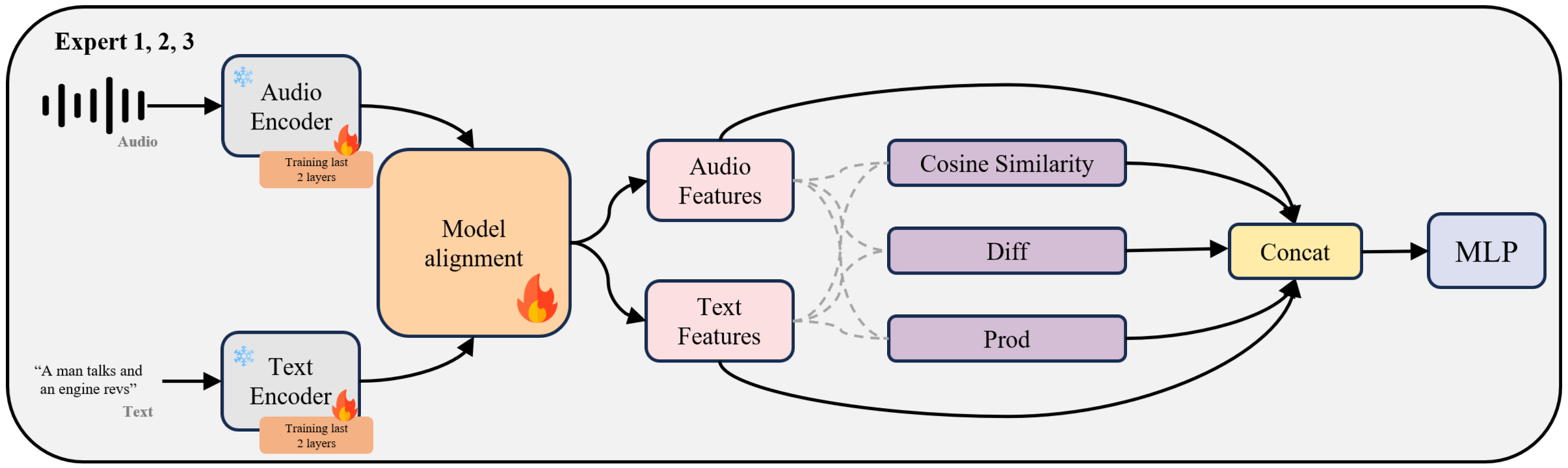}}
  \centerline{(c) The Fine-Tuned Expert-1, Expert-2, Expert-3 Architecture}\medskip
\end{minipage}
\vspace{-9mm}
\caption{Architecture of the proposed text-audio relevance predictor.}
\label{fig:architecture}
\vspace{-6mm}
\end{figure}

As shown in Fig ~\ref{fig:architecture}, we employ a Mixture of Experts (MoE) architecture to leverage the complementary strengths of multiple specialized models for the task of predicting the TARS of a synthesized audio-text pair. The MoE framework dynamically combines the predictions of several expert models through a gating mechanism, allowing for a more nuanced and accurate estimation.
Specifically, our system comprises \textbf{four specialized expert models} with different architectures and training strategies.
For \textbf{Experts 1–3}, we employ pre-trained CLAP-based~\cite{laionclap2023} models and generate the TARS by computing the similarity between audio and text features. For \textbf{Expert 4}, we incorporate a SeqCoAttn module to enable more fine-grained audio-text alignment.

\textbf{Expert 1} uses the LAION-CLAP ~\cite{laionclap2023} backbone, a dual-encoder model trained via large-scale audio-text contrastive learning. This pre-trained model projects audio and text into a shared latent space, excelling at zero-shot audio classification and text-audio retrieval. In the MoE framework, it provides generalizable cross-modal semantic cues to complement other experts. 

\textbf{Expert 2} is based on MGA-CLAP ~\cite{li2024advancing}, a CLAP extension that improves multi-grained audio–text alignment through guided attention mechanisms. By enhancing interactions between local audio patterns and global semantic representations, MGA-CLAP provides stronger fine-grained semantic correspondence in complex audio scenes, complementing the global alignment strength of CLAP.

\textbf{Expert 3} builds on M2D-CLAP ~\cite{niizumi2025m2d-clap}, a multi-stage framework that integrates self-supervised audio representation learning with CLAP-style cross-modal contrastive learning. By learning general-purpose audio features before semantic alignment with text embeddings, M2D-CLAP produces robust audio–language representations that generalize well beyond standard CLAP settings, supporting reliable semantic scoring and regression-style evaluation.

\textbf{Expert 4} is developed based on the baseline system, integrating a BEATs audio encoder ~\cite{Chen2022beats}, RoBERTa text encoder ~\cite{liu2019roberta}, and SeqCoAttn fusion module to model cross-modal temporal correspondence. Audio layer-wise features are aggregated with learnable weights and projected to a dimension of 512; text embeddings are aligned to the same dimension. Bidirectional SeqCoAttn enables cross-modality attention, followed by adaptive max-pooling and a two-layer MLP (with dropout/Tanh clipping) to output alignment scores. It specializes in local temporal matching, complementing the strengths of Experts 1–3.

All four expert models were individually fine-tuned on the official training set. During MoE training, the expert parameters were frozen, and only the gating network was optimized to effectively integrate the experts.

\subsection{SeqCoAttn}
SeqCoAttn employs bidirectional cross-attention between audio and text sequences using two multi-head attention layers.
The attended sequences are pooled via adaptive max-pooling and concatenated to form a fused representation.

\subsection{Loss function}

We adopt the baseline’s loss formulation~\cite{kanamori_INTERSPEECH2025}, which combines a clipped Mean Squared Error (MSE) loss and a margin-based contrastive loss. The contrastive loss is defined as: 

\vspace{-2mm}
\begin{equation}
    \mathcal{L}_{\text{con}}=\max(0, \left | \hat{d} -d \right | - \varepsilon )
\end{equation}
\vspace{-4mm}

where $\hat{d}$ and $d$ denote the predicted and ground-truth score differences between sample pairs. The clipped MSE loss is given by: 

\vspace{-5mm}
\begin{equation}
\mathcal{L}_{\text{ClippedMSE}}=\frac{1}{N}\sum_{i=1}^{N} \mathcal{I}  (\left | \hat{y}-y_{i} \right |>\tau) \cdot (\hat{y}-y_{i})^{2}
\end{equation}
\vspace{-4mm}

where $y_{i} $ and $\hat{y} _{i} $ denote the ground-truth and predicted scores for the i-th utterance, respectively. The final loss function is defined as: \(\mathcal{L=\beta \cdot L_{\text{mse}} + \gamma \cdot L_{\text{con}} } \)  where $\beta$ and $\gamma$ are hyperparameters.


\section{RESULTs and discussions}
The model’s performance on the official blind test set is summarized in Table~\ref{tab:result}. As shown, our proposed method consistently outperforms the baseline across all evaluation metrics, achieving an absolute improvement of 30.6\% in SRCC.

To further analyze the contribution of individual experts and the effect of expert aggregation, we conduct a detailed ablation study on the validation set, as reported in Table~\ref{tab:ablation result}. Experts 1–3 achieve comparable performance owing to their strong CLAP-based cross-modal representations, while Expert 4 (SeqCoAttn-based expert), which emphasizes sequential cross-attention and temporal correspondence, exhibits lower standalone performance. Nevertheless, aggregating multiple experts through the MoE framework consistently improves all evaluation metrics. The MoE model with three experts (Experts 1-3) already outperforms all individual experts, indicating effective fusion of complementary semantic cues. Further improvements are observed when incorporating all four experts, with the MoE (4 Experts) achieving the best SRCC, LCC, and KTAU, as well as the lowest MSE. These results demonstrate that expert diversity and dynamic gating enable the model to capture both global semantic alignment and fine-grained temporal correspondence, validating the effectiveness and scalability of the proposed MoE architecture.

\begin{table}[t!]
    \centering
    \vspace{-5mm}
        \caption{Performance comparison between our model and the baseline on the official test set.}
        \vspace{1mm}
    \begin{tabular}{lcccc}
    \toprule
        \textbf{System} & \textbf{SRCC $\uparrow$} & \textbf{LCC $\uparrow$} & \textbf{KTAU $\uparrow$} & \textbf{MSE $\downarrow$} \\
        \midrule
        Ours &  \textbf{0.6402} & \textbf{0.6873} & \textbf{0.4612} & \textbf{3.0111} \\
        Baseline & 0.3345 & 0.3420 & 0.2290 & 4.8110 \\
        \bottomrule
    \end{tabular}

    \label{tab:result}
\end{table}

\begin{table}[t!]
    \centering
        \caption{Performance comparison of individual experts and MoE variants on the official validation set.}
        \vspace{1mm}
    \begin{tabular}{lcccc}
    \toprule
        \textbf{System} & \textbf{SRCC $\uparrow$} & \textbf{LCC $\uparrow$} & \textbf{KTAU $\uparrow$} & \textbf{MSE $\downarrow$} \\
        \midrule
        Expert 1 &  {0.6302} & {0.6492} & {0.4542} & {3.5382} \\
        Expert 2 &  {0.6297} & {0.6562} & {0.4550} & {3.7208} \\
        Expert 3 &  {0.6257} & {0.6438} & {0.4496} & {3.7369} \\
        Expert 4 &  {0.5808} & {0.5865} & {0.4147} & {3.8082} \\
        \midrule
        MoE (3 Experts) &  {0.6480} & {0.6745} & {0.4693} & {3.3592} \\
        MoE (4 Experts) &  \textbf{{0.6680} }& \textbf{{0.6845}} & \textbf{{0.4861}} & \textbf{{3.3462}} \\
        \bottomrule
    \end{tabular}

    \label{tab:ablation result}
\end{table}




\section{CONCLUSION}

To tackle the inefficiency and inconsistency of subjective evaluation for text-to-audio (TTA) semantic alignment, we propose an objective evaluator based on a Mixture of Experts (MoE) and Sequential Cross-Attention (SeqCoAttn).
We employ four specialized experts, which are dynamically fused via a gating network, and a hybrid loss boosts robustness and ranking accuracy. On the official blind test dataset, our model outperforms the baselines by 30.6\% in SRCC and ranks first, demonstrating strong alignment with human judgment. Future work will further analyze expert selection behavior and extend the proposed framework beyond the challenge setting.

\balance
\bibliographystyle{IEEEbib}
\bibliography{strings,refs}

\begin{thebibliography}{10}

\bibitem{liu2023audioldm}
Haohe Liu, Zehua Chen, Yi~Yuan, Xinhao Mei, Xubo Liu, Danilo Mandic, Wenwu Wang, and Mark~D Plumbley,
\newblock ``Audioldm: Text-to-audio generation with latent diffusion models,''
\newblock {\em arXiv preprint arXiv:2301.12503}, 2023.

\bibitem{liu2024audiolcm}
Huadai Liu, Rongjie Huang, Yang Liu, Hengyuan Cao, Jialei Wang, Xize Cheng, Siqi Zheng, and Zhou Zhao,
\newblock ``Audiolcm: Text-to-audio generation with latent consistency models,''
\newblock {\em arXiv preprint arXiv:2406.00356}, 2024.

\bibitem{huang2023make2}
Jiawei Huang, Yi~Ren, Rongjie Huang, Dongchao Yang, Zhenhui Ye, Chen Zhang, Jinglin Liu, Xiang Yin, Zejun Ma, and Zhou Zhao,
\newblock ``Make-an-audio 2: Temporal-enhanced text-to-audio generation,''
\newblock {\em arXiv preprint arXiv:2305.18474}, 2023.

\bibitem{ren2019fastspeech}
Yi~Ren, Yangjun Ruan, Xu~Tan, Tao Qin, Sheng Zhao, Zhou Zhao, and Tie-Yan Liu,
\newblock ``Fastspeech: Fast, robust and controllable text to speech,''
\newblock {\em Advances in neural information processing systems}, vol. 32, 2019.

\bibitem{arik2017deep}
Sercan~{\"O} Ar{\i}k, Mike Chrzanowski, Adam Coates, Gregory Diamos, Andrew Gibiansky, Yongguo Kang, Xian Li, John Miller, Andrew Ng, Jonathan Raiman, et~al.,
\newblock ``Deep voice: Real-time neural text-to-speech,''
\newblock in {\em International conference on machine learning}. PMLR, 2017, pp. 195--204.

\bibitem{huang2023make}
Rongjie Huang, Jiawei Huang, Dongchao Yang, Yi~Ren, Luping Liu, Mingze Li, Zhenhui Ye, Jinglin Liu, Xiang Yin, and Zhou Zhao,
\newblock ``Make-an-audio: Text-to-audio generation with prompt-enhanced diffusion models,''
\newblock in {\em International Conference on Machine Learning}. PMLR, 2023, pp. 13916--13932.

\bibitem{1}
Dongchao Yang, Jianwei Yu, Helin Wang, Wen Wang, Chao Weng, Yuexian Zou, and Dong Yu,
\newblock ``Diffsound: Discrete diffusion model for text-to-sound generation,''
\newblock {\em IEEE/ACM Transactions on Audio, Speech, and Language Processing}, vol. 31, pp. 1720--1733, 2023.

\bibitem{yin2025envsdd}
Han Yin, Yang Xiao, Rohan~Kumar Das, Jisheng Bai, Haohe Liu, Wenwu Wang, and Mark~D Plumbley,
\newblock ``Envsdd: Benchmarking environmental sound deepfake detection,''
\newblock {\em arXiv preprint arXiv:2505.19203}, 2025.

\bibitem{wang2025tta}
Hui Wang, Cheng Liu, Junyang Chen, Haoze Liu, Yuhang Jia, Shiwan Zhao, Jiaming Zhou, Haoqin Sun, Hui Bu, and Yong Qin,
\newblock ``Tta-bench: A comprehensive benchmark for evaluating text-to-audio models,''
\newblock {\em arXiv preprint arXiv:2509.02398}, 2025.

\bibitem{lee2024challenge}
Junwon Lee, Modan Tailleur, Laurie~M Heller, Keunwoo Choi, Mathieu Lagrange, Brian McFee, Keisuke Imoto, and Yuki Okamoto,
\newblock ``Challenge on sound scene synthesis: Evaluating text-to-audio generation,''
\newblock {\em arXiv preprint arXiv:2410.17589}, 2024.

\bibitem{masoudnia2014mixture}
Saeed Masoudnia and Reza Ebrahimpour,
\newblock ``Mixture of experts: a literature survey,''
\newblock {\em Artificial Intelligence Review}, vol. 42, no. 2, pp. 275--293, 2014.

\bibitem{laionclap2023}
Yusong Wu*, Ke~Chen*, Tianyu Zhang*, Yuchen Hui*, Taylor Berg-Kirkpatrick, and Shlomo Dubnov,
\newblock ``Large-scale contrastive language-audio pretraining with feature fusion and keyword-to-caption augmentation,''
\newblock in {\em IEEE International Conference on Acoustics, Speech and Signal Processing, ICASSP}, 2023.

\bibitem{li2024advancing}
Yiming Li, Zhifang Guo, Xiangdong Wang, and Hong Liu,
\newblock ``Advancing multi-grained alignment for contrastive language-audio pre-training,''
\newblock in {\em Proceedings of the 32nd ACM International Conference on Multimedia}, 2024, pp. 7356--7365.

\bibitem{niizumi2025m2d-clap}
Daisuke Niizumi, Daiki Takeuchi, Masahiro Yasuda, Binh Thien~Nguyen, Yasunori Ohishi, and Noboru Harada,
\newblock ``M2d-clap: Exploring general-purpose audio-language representations beyond clap,''
\newblock {\em IEEE Access}, vol. 13, pp. 163313--163330, 2025.

\bibitem{Chen2022beats}
Sanyuan Chen, Yu~Wu, Chengyi Wang, Shujie Liu, Daniel Tompkins, Zhuo Chen, and Furu Wei,
\newblock ``Beats: Audio pre-training with acoustic tokenizers,''
\newblock 2022.

\bibitem{liu2019roberta}
Yinhan Liu, Myle Ott, Naman Goyal, Jingfei Du, Mandar Joshi, Danqi Chen, Omer Levy, Mike Lewis, Luke Zettlemoyer, and Veselin Stoyanov,
\newblock ``Roberta: A robustly optimized bert pretraining approach,''
\newblock {\em arXiv preprint arXiv:1907.11692}, 2019.

\bibitem{kanamori_INTERSPEECH2025}
Yusuke Kanamori, Yuki Okamoto, Taisei Takano, Shinnosuke Takamichi, Yuki Saito, and Hiroshi Saruwatari,
\newblock ``Relate: Subjective evaluation dataset for automatic evaluation of relevance between text and audio,''
\newblock in {\em Proc. INTERSPEECH}, 2025.

\end{thebibliography}

\end{document}